\documentclass[%
 aip,
 sd,%
 amsmath,amssymb,
reprint,%
]{revtex4-1}

\usepackage{graphicx}% Include figure files
\usepackage{dcolumn}% Align table columns on decimal point
\usepackage{bm}% bold math
\usepackage{algorithm}
\usepackage[noend]{algpseudocode}

\begin{document}

\preprint{AIP/123-QED}

\title[R. Zhang and S. Pei]{Dynamic range maximization in excitable networks}% Force line breaks with \\

\author{Renquan Zhang}
\email{zhangrenquan@dlut.edu.cn}
\affiliation{School of Mathematical Sciences, Dalian University of Technology, Dalian 116024, China}

\author{Sen Pei}
\email{sp3449@cumc.columbia.edu}
\affiliation{Department of Environmental Health Sciences, Mailman School of Public Health, Columbia University, New York, NY 10032, USA}

\date{\today}% It is always \today, today,

\begin{abstract}
We study the strategy to optimally maximize the dynamic range of excitable networks by removing the minimal number of links. A network of excitable elements can distinguish a broad range of stimulus intensities and has its dynamic range maximized at criticality. In this study, we formulate the activation propagation in excitable networks as a message passing process in which the critical state is reached when the largest eigenvalue of the weighted non-backtracking (WNB) matrix is exactly one. By considering the impact of single link removal on the largest eigenvalue, we develop an efficient algorithm that aims to identify the optimal set of links whose removal will drive the system to the critical state. Comparisons with other competing heuristics on both synthetic and real-world networks indicate that the proposed method can maximize the dynamic range by removing the smallest number of links, and at the same time maintain the largest size of the giant connected component.
\end{abstract}

\maketitle

\begin{quotation}
Networks of coupled excitable elements can properly describe many real-world systems. The nonlinear collective dynamics of an excitable network allow the system to distinguish stimulus intensities spanning several orders of magnitude, manifested by a broad dynamic range. In practice, how to maximize the dynamic range under the minimal control of the network structure has profound implications in designing excitable systems that can achieve optimal functionality. In this work, we show that the dynamic range is controlled by the largest eigenvalue of the weighted non-backtracking (WNB) matrix. Based on the stability analysis of the WNB matrix, we propose a heuristic algorithm that outperforms other widely used ranking methods in dynamic range maximization. This study provides a practical method to locate high-impact links in general dynamical processes governed by the largest eigenvalue of the WNB matrix.
\end{quotation}

\section{\label{sec:1}Introduction}

A defining characteristic of sensory systems is their ability to distinguish intensities of physical stimuli varied by a wide range of magnitude \cite{cleland1999concentration,copelli2014criticality}. Despite that single neurons usually respond to external stimuli in a linear way \cite{reisert2001response}, it was revealed that such nonlinear transduction of information could be realized by the collective dynamics of coupled excitable elements \cite{kinouchi2006optimal,copelli2002physics,copelli2005intensity}. As a result, the excitable network has become a key model in the field of psychophysics \cite{teghtsoonian1975psychophysics}. In a broader context, many biological, social and engineering systems can be properly modeled by networks of interacting excitable nodes. So far, the excitable network model has been used in a variety of applications, for instance, information processing in sensory neural networks \cite{larremore2014inhibition,copelli2005signal,gollo2016diversity}, signal transmission in brain networks \cite{gautam2015maximizing,williams2014quasicritical,marro2013signal}, and influence detection in social networks \cite{pei2015detecting}.

The central question in the study of excitable networks is how they respond to external stimuli whose intensities usually vary by several orders of magnitude. The performance of an excitable network can be evaluated by the dynamic range, which quantifies the range of stimulus intensities that can be robustly distinguished by the system. In particular, the dynamic range of an excitable network is determined by the topology of the network connections: the optimal (or maximal) dynamic range is achieved at the critical state of the system \cite{kinouchi2006optimal,larremore2011predicting,larremore2011effects}. In homogeneous random graphs, Kinouchi and Copelli found that the critical state occurred when the branching ratio $\sigma=1$ \cite{kinouchi2006optimal}. For more general network structure, Larremore {\it et al.} found that the critical state was characterized by the unit largest eigenvalue of the weighted adjacency matrix $\lambda_{W}=1$ \cite{larremore2011predicting}.

In applications, a higher dynamic range of excitable systems implies a better functionality in detecting external stimuli. Therefore, it is desirable to maximize the dynamic range without changing the network structure dramatically. Many previous studies have explored the strategy to optimize the dynamic range \cite{gautam2015maximizing,pei2012enhance,gollo2009active}. In this work, we study how to perturb the excitable networks to criticality via the deletion of a minimal number of connections. To do this, we first formulate the activation propagation dynamics in excitable networks as a message passing process, in which the focal nodes are assumed to be ``virtually'' removed from the network \cite{mezard2009information}. By linearizing the system about the zero solution, the message passing equations recover the same form for the bond percolation process \cite{karrer2014percolation,hamilton2014tight}. The critical state is therefore governed by the largest eigenvalue of the weighted non-backtracking (WNB) matrix $\lambda_{NB}$, which controls the stability of the zero solution. Through simulations on various networks, we verify that the dynamic range is maximized when the largest eigenvalue of the WNB matrix is exactly one, i.e., $\lambda_{NB}=1$. %As we will show later, this result is consistent with the finding on weighted adjacency matrix of sparse directed networks discussed in Larremore {\it et al.} \cite{?}.

Based on the linearized system, we explore the impact of link removal on the largest eigenvalue of the WNB matrix. With proper approximations, we derive the analytical form of the collective influence (CI) of each link on the largest eigenvalue of the WNB matrix \cite{morone2015influence}. By adaptively removing the links with high CI scores, we propose a heuristic algorithm to find the optimal set of influential links whose removal would drive the system to the critical state. We compare the performance of the proposed algorithm with several commonly used link ranking methods, including weight-based, degree-based and eigenvector-based ranking algorithms. Results indicate that the collective influence algorithm outperforms other competing methods in three aspects: 1) the number of links removed to maximize the dynamic range is smaller; 2) the dynamic range at the critical state is higher; and 3) the size of the largest connected component during the link removal remains larger. Because the collective influence of links only depends on the form of the WNB matrix, this algorithm can be adapted and used for identifying influential links in other dynamical processes whose dynamics are controlled by the largest eigenvalue of the WNB matrix.

\section{\label{sec:2}Dynamic range of excitable networks}

The dynamical evolution of nodes' states in an excitable network can be described by the Kinouchi-Copelli model \cite{kinouchi2006optimal}. Considering a system composed on $N$ nodes and $2M$ directed links, each node $i$ can be in one of $m+1$ states: the resting state $x_i=0$, the excited state $x_i=1$, and the refractory states $x_i=2,3,\cdots,m$ ($m\geq2$). At each time step $t$, a resting node $i$ ($x_i^t=0$) can be excited by its neighbor $j$ in excited state ($x_j^t=1$) with probability $a_{ij}$, or independently by an external stimulus with probability $\eta$. Meanwhile, nodes in excited or refractory state ($x_i^t\geq 1$) will transit to the next refractory state deterministically, until $x_i^t=m$ which will return to the resting state. That is, $x_i^{t+1}=x_i^t+1$ if $1\leq x_i^t<m$ and $x_i^{t+1}=0$ if $x_i^t=m$. In the excitable network, the topology of connections and pairwise strength of interactions are described by the weighted adjacency matrix $W=\{a_{ij}\}_{N\times N}$.

For a given external stimulus strength $\eta$, the network response $F$ is defined as
\begin{equation}\label{response}
F=\left\langle f\right\rangle_{t}=\lim_{T\to\infty}\frac{1}{T}\sum_{t=0}^Tf^t,
\end{equation}
where $\left\langle \cdot\right\rangle_{t}$ denotes an average over time and $f^{t}$ is the fraction of excited nodes at time $t$. With a larger stimulus strength $\eta\in(0,1)$, we have a larger value of response $F$. Therefore, $F=F(\eta)$ is a monotonically increasing function of $\eta$. In excitable networks, the dynamic range $\Delta$ quantifies the range of external stimuli that result in distinguishable network responses $F$. To discard the responses that are negligible or close to saturation, we select a lower stimulus threshold $\eta_{low}$ and an upper stimulus threshold $\eta_{high}$. The dynamic range is then defined as
\begin{eqnarray}\label{Delta}
\Delta=10\log_{10}\frac{\eta_{high}}{\eta_{low}}.
\end{eqnarray}
In this work, we select the lower threshold $\eta_{0.1}$ (upper threshold $\eta_{0.9}$) to exclude the lower $10\%$ (upper $10\%$) network responses. In particular, $\eta_{0.1}$ and $\eta_{0.9}$ are determined from their corresponding responses $F_{0.1}$ and $F_{0.9}$, where $F_x=F_0+x(F_{max}-F_0)$ (Here $F_0=\lim_{\eta\to0}F$ and $F_{max}$ are the minimum and maximum responses of the network).

\section{\label{sec:3}Criticality and non-backtracking matrix}

We denote the probability that a given node $i$ is excited at time $t$ by $p_i^t$. For a directed link from $i$ to $j$ ($i\to j$), suppose node $j$ is ``virtually'' removed from the network (i.e., creating a ``cavity'' at node $j$) and reconsider if node $i$ is excited or not \cite{morone2015influence}. This information can be stored in an auxiliary variable $p_{i\to j}^t$ representing the probability of node $i$ being excited in the absence of node $j$ at time $t$. For $m=1$, the system has only two states, resting and excited. For {\it sparse} networks, the updating process can be described by
\begin{equation}\label{pij_m=1}
p_{i\to j}^{t+1}=(1-p_{i\to j}^{t})( \eta+(1-\eta)[1-\prod_{k\setminus j}( 1-p_{k\to i}^{t}a_{ik}A_{ik})]),
\end{equation}
where $A_{ij}$ is the element of the binary adjacency matrix ($A_{ij}=1$ if $a_{ij}>0$, and $A_{ij}=0$ otherwise) and $k\setminus j$ means $k$ runs over all nodes except $j$. Putting back $j$ into the network, we can get the probability of node $i$ being excited at time $t+1$ through
\begin{equation}\label{pi_m=1}
p_{i}^{t+1}=(1-p_{i}^{t})( \eta+( 1-\eta)[1-\prod_{k}(1-p_{k\to i}^{t}a_{ik}A_{ik})]).
\end{equation}

For general situations with $m\geq 2$, the probability of node $i$ being at any refractory states in the absence of node $j$ at time $t$ is $\sum_{l=1}^{m-1}p_{i\to j}^{t-l}$, so the probability of node $i$ being at the resting state is $1-\sum_{l=0}^{m-1}p_{i\to j}^{t-l}$. The updating process therefore follows:
\begin{equation}\label{pij_any_m}
p_{i\to j}^{t+1}=(1-\sum_{l=0}^{m-1}p_{i\to j}^{t-l})( \eta+( 1-\eta)[1-\prod_{k\backslash j}( 1-p_{k\to i}^{t}a_{ik}A_{ik}) ] ),
\end{equation}
\begin{equation}\label{pi_any_m}
p_{i}^{t+1}=(1-\sum_{l=0}^{m-1}p_{i}^{t-l})( \eta+( 1-\eta)[1-\prod_{k}( 1-p_{k\to i}^{t}a_{ik}A_{ik})]).
\end{equation}
Denote $\lim_{t\to\infty}p_{i\to j}^t=p_{i\to j}$ and $\lim_{t\to\infty}p_{i}^t=p_{i}$. Equations (\ref{pij_any_m}-\ref{pi_any_m}) in the steady state become
\begin{equation}\label{pij}
p_{i\to j}=(1-mp_{i\to j})( \eta+( 1-\eta)[1-\prod_{k\backslash j}( 1-p_{k\to i}a_{ik}A_{ik})])
\end{equation}
and
\begin{equation}\label{pi}
p_{i}=(1-mp_{i})( \eta+( 1-\eta)[1-\prod_{k}( 1-p_{k\to i}a_{ik}A_{ik})]).
\end{equation}

To solve above self consistent equations, let us denote $G_{i\to j}(\eta,p_{\to})= \eta+( 1-\eta)[1-\prod_{k\backslash j}( 1-p_{k\to i}a_{ik}A_{ik})] $ and $G_{i}(\eta,p_{\to})= \eta+( 1-\eta)[1-\prod_{k}( 1-p_{k\to i}a_{ik}A_{ik})]$. After rearranging the equations, we have
\begin{equation}\label{pij_G}
p_{i\to j}=\frac{G_{i\to j}(\eta,p_{\to})}{mG_{i\to j}(\eta,p_{\to})+1}
\end{equation}
and
\begin{equation}\label{pi_G}
p_{i}=\frac{G_{i}(\eta,p_{\to})}{mG_{i}(\eta,p_{\to})+1}.
\end{equation}
Notice that $p_{i\to j}$ and $p_{i}$ are actually the functions of $\eta$. As a result, the response $F=F(\eta)$ can be written as follows:
\begin{equation}\label{F}
F(\eta)=\lim_{T\to\infty}\frac{1}{T}\sum_{t=0}^{T}\frac{1}{N}\sum_{i=1}^{N}p_{i}^{t}(\eta)=\frac{1}{N}\sum_{i=1}^{N}p_{i}(\eta).
\end{equation}

For $\eta=1$, it is trivial that $G_{i\to j}(1,p_{\to})=1$ and $G_{i}(1,p_{\to})=1$. This leads to a constant maximal response $F_{max}=1/ (m+1)$ regardless of the network structure. The dynamic range thus only depends on the minimal response $F_0=\lim_{\eta\to0}F(\eta)$. When $\eta=0$, Eq. (\ref{pij_G}) always has a trivial solution  $\{ p_{i\to j}=0 \}$ for all $i\to j$, which in turn gives $\{p_{i}=0\}$ for all node $i$ and thus $F_{0}=0$. The stability of this solution is controlled by the largest eigenvalue of the linear operator represented by the $2M\times2M$ matrix defined on the directed links $k\to l, i\to j$:
\begin{eqnarray}\label{partial_pij}
\mathcal{M}_{k\to l,i\to j}&=&\frac{\partial p_{i\to j}}{\partial p_{k\to l}}\bigg\vert_{\{\eta=0, p_{i\to j}=0 \}} \nonumber \\
&=&\frac{\frac{\partial G_{i\to j}}{\partial p_{k\to l}}(mG_{i\to j}+1)-mG_{i\to j}\frac{\partial G_{i\to j}}{\partial p_{k\to l}}}{(mG_{i\to j}+1)^2}\nonumber \\
&=&\frac{\partial G_{i\to j}}{\partial p_{k\to l}}\bigg\vert_{\{\eta=0, p_{i\to j}=0 \}}.
\end{eqnarray}
Here $G_{i\to j}=0$ when $\eta=0$ and $\{ p_{i\to j}=0 \}$ for all $i\to j$. According to the definition of $G_{i\to j}$, the partial derivative of $G_{i\to j}$ is given by
\begin{equation}\label{partial_G}
\frac{\partial G_{i\to j}}{\partial p_{k\to l}}\bigg\vert_{\{\eta=0, p_{i\to j}=0 \}}=\bigg\{
\begin{array}{cc}
a_{lk}A_{lk}  &\text{if $l=i$ and $j\neq k$,}\\
0   &\text{otherwise.} \\
\end{array}
\end{equation}
Going back to Eq.(\ref{partial_pij}), we obtain the $2M\times2M$ matrix $\widehat{\mathcal{M}}=\{\mathcal{M}_{k\to l,i\to j}\}$:
\begin{eqnarray}\label{NB}
\mathcal{M}_{k\to l,i\to j}=\bigg\{
\begin{array}{cc}
a_{lk}A_{lk}  &\text{if $l=i$ and $j\neq k$,} \\
0   &\text{otherwise.} \\
\end{array}
\end{eqnarray}
The matrix $\widehat{\mathcal{M}}$ is a generalization of the non-backtracking (NB) matrix of networks \cite{hashimoto1989zeta}, which recently has found many applications in network science \cite{morone2015influence,martin2014localization,radicchi2016leveraging,karrer2014percolation,hamilton2014tight}. Therefore, we refer $\widehat{\mathcal{M}}$ as the weighted non-backtracking (WNB) matrix of the excitable network. The stability of the zero solution is determined by the largest eigenvalue of the WNB matrix $\lambda_{NB}=1$: stable if $\lambda_{NB}<1$ and unstable if $\lambda_{NB}>1$.

In the subcritical regime with $\lambda_{NB}<1$, the minimal response $F_0$ remains zeros. As $\lambda_{NB}$ increases, the sensitivity of the network is enlarged because the amplifying effect on weak stimuli caused by activity propagation among neighbors becomes stronger. As a result, the dynamic range $\Delta$ increases monotonically with $\lambda_{NB}$. In the supercritical regime with $\lambda_{NB}>1$, the minimal response $F_0$ becomes nonzero and masks the existence of weak stimuli, which reduces the dynamic range $\Delta$. Consequently, the maximal dynamic range is achieved at the critical state where $\lambda_{NB}=1$ \cite{kinouchi2006optimal}.

\begin{figure}
\includegraphics[width=1\columnwidth]{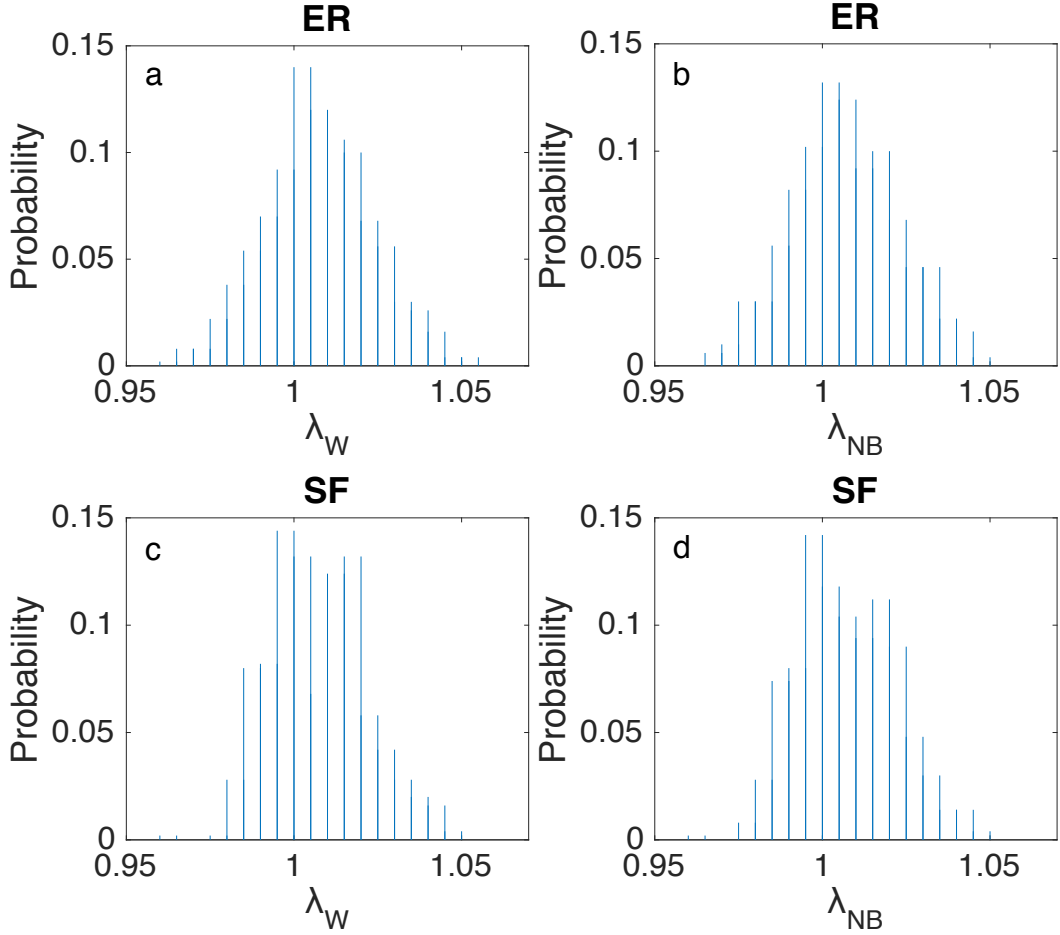}
\caption{\label{threshold} For directed ER networks ($N=5000$, $K=2.5$), the distributions of the largest eigenvalues of the weighted adjacency matrix $\lambda_{W}$ (a) and the WNB matrix $\lambda_{NB}$ (b) when the dynamic range is maximized are displayed. Results are obtained from 1000 independent realizations. Same analyses for directed SF networks ($N=5000$, $\gamma=3$) are presented in (c) and (d).}
\end{figure}

To verify this theoretical result, we performed simulations on synthetic random networks. In particular, we are interested in the largest eigenvalue of the WNB matrix $\lambda_{NB}$ when the dynamic range is maximized. In previous study, it was found that the optimal dynamic range appears if the largest eigenvalue of the weighted adjacency matrix $\lambda_{W}$ is one \cite{larremore2011predicting}. To compare these results, in our simulations, we also reported the values of $\lambda_{W}$ when the maximal dynamic range was observed.

We start from homogeneous random networks. To generate directed Erd\"{o}s-R\'{e}nyi (ER) networks, we first generated independent in-degree and out-degree sequences from a Poisson degree distribution $P(k)=K^ke^{-K}/k!$ where $K$ is the average degree. Then we connected the nodes using the configuration model \cite{molloy1995critical}. For simplicity, we assume the weights of all links are constant. By varying the link weight, we can control both $\lambda_{NB}$ and $\lambda_{W}$. In simulations, we generated 1000 independent realizations of the directed ER networks with network size $N=5000$ and average degree $K=2.5$. For each network, we slowly tuned up the link weight, calculated the dynamic range and recorded $\lambda_{NB}$ and $\lambda_{W}$ when $\Delta$ reached the peak. The distributions of the obtained $\lambda_{NB}$ and $\lambda_{W}$ are shown in Fig. \ref{threshold}(a-b). We further performed the same analysis on directed scale-free (SF) networks, generated using a power-law degree distribution $P(k)\sim k^{-\gamma}$ and the configuration model ($N=5000$, $\gamma=3$). Specifically, we first generated in-degree and out-degree sequences from the distribution $P(k)\sim k^{-\gamma}$ with $k_{min}=2$ and $k_{max}=1000$, and then connected the ``stubs'' randomly. Results are presented in Fig. \ref{threshold}(c-d). The narrow distributions tightly surrounding $\lambda_{NB}=1$ and $\lambda_{W}=1$ indicate that, for spare random networks, 1) the dynamic range is maximized when the largest eigenvalue of the WNB matrix is exactly one; and 2) this finding is consistent with previous result on weighted adjacency matrix.

The above result is reminiscent of recent findings on bond percolation \cite{karrer2014percolation,hamilton2014tight}. On sparse networks, it was found the percolation threshold is given by the inverse of the largest eigenvalue of the non-backtracking (NB) matrix. This result is equivalent to previous finding that, for sparse networks, the critical occupation probability for percolation is the reciprocal of the largest eigenvalue of the adjacency matrix \cite{bollobas2010percolation}. For dense networks, both estimations give a lower bound of the real threshold value, but the estimation based on NB matrix is tighter \cite{hamilton2014tight}.

\section{\label{sec:4}Dynamic range maximization}

In applications, it is common to encounter the situation where we want to modify an existing excitable network to maximize its dynamic range with a lower cost. Given that many different ways of modifications may exist, in this work, we assume only the network structure can be changed while the link weights remain fixed. Because the dynamic range is optimized at the critical state, the problem becomes to adjust the connections in the network so that the largest eigenvalue of the WNB matrix $\lambda_{NB}$ is one. Here we considering the following optimization problem: for an excitable network with $\lambda_{NB}>1$, find the minimal set of links whose removal will lead to $\lambda_{NB}=1$. For networks with $\lambda_{NB}<1$, the adjustment may require addition of new links. We leave this problem to future works.

To find the optimal link set, we first define the {\it collective influence} (CI) of each link on the largest eigenvalue $\lambda_{NB}$. The collective influence theory was developed to find multiple influential nodes in optimal percolation \cite{morone2015influence}, and has been applied to independent spreading model \cite{teng2016collective} and linear threshold model \cite{sen2017efficient}. Here we generalize it to work for network links. The main idea is to quantify the impact of each link's removal on the value of $\lambda_{NB}$, which is collectively determined by the global network structure.

We use the power method to estimate the largest eigenvalue of the WNB matrix \cite{morone2015influence}. For a $2M$ dimensional nonzero vector $\textbf{w}_{0}$, denote
\begin{eqnarray}\label{PM}
\textbf{w}_{\ell}=\widehat{\mathcal{M}}^{\ell}\textbf{w}_{0}.
\end{eqnarray}
The largest eigenvalue of $\widehat{\mathcal{M}}$ is
\begin{eqnarray}\label{lambda}
\lambda_{NB}=\lim_{\ell\to\infty}\bigg\lbrack\frac{\left| \textbf{w}_{\ell}\right| }{ \left| \textbf{w}_{0}\right| }\bigg\rbrack^{1/\ell},
\end{eqnarray}
where
\begin{eqnarray}\label{module}
\left| \textbf{w}_{\ell}\right|^{2}=\langle \textbf{w}_{\ell} \vert \textbf{w}_{\ell}\rangle=\langle \textbf{w}_{0}\vert ( \widehat{\mathcal{M}}^{\ell}) ^{T}	\widehat{\mathcal{M}}^{\ell} \vert \textbf{w}_{0} \rangle.
\end{eqnarray}
For notation convenience, we put $\mathcal{M}_{k\to l,i\to j}$ into a larger space of dimension $N\times N\times N\times N$:
\begin{eqnarray}\label{NB_E}
\widehat{\mathcal{M}}_{klij}=a_{lk}A_{lk}A_{ji}\delta_{li}\left( 1-\delta_{kj} \right).
\end{eqnarray}
Here $\delta_{ij}=1$ if and only if $i=j$. Supposing the initial vector is $\textbf{w}_{0}=\left\lbrace \vert w_{1}\rangle_{ij} \right\rbrace =\left\lbrace A_{ji}\right\rbrace$, we can calculate the iteration starting from $\ell=1$:
\begin{eqnarray}\label{w1ij_left}
{}_{ij}\langle w_{1}\vert&=&\sum_{kl} {}_{kl}\langle w_{0}\vert\widehat{\mathcal{M}}_{klij}\nonumber \\
&=&\sum_{kl}a_{lk}A_{lk}A_{ji}\delta_{li}\left( 1-\delta_{kj} \right)\nonumber \\
&=&A_{ji}\sum_{k\backslash j}a_{ik}A_{ik},
\end{eqnarray}
\begin{eqnarray}\label{w1ji_right}
\vert w_{1}\rangle_{ij}&=&\sum_{kl} \widehat{\mathcal{M}}_{ijkl}\vert w_{0}\rangle_{kl}\nonumber \\
&=&\sum_{kl}a_{ji}A_{ji}A_{lk}\delta_{jk}\left( 1-\delta_{il} \right)\nonumber \\
&=&a_{ji}A_{ji}\sum_{l\backslash i}A_{lj}.
\end{eqnarray}

Define $S_{i\backslash j}^{in}=\sum_{k\backslash j}a_{ik}A_{ik}$ and $K_{j\backslash i}^{out}=\sum_{l\backslash i}A_{lj}$. We have
\begin{eqnarray}\label{lamda1}
\left| \textbf{w}_{1}\right|^{2}&=&\sum_{ij}{}_{ij}\langle w_{1} \vert w_{1}\rangle_{ij}\nonumber \\
&=&\sum_{ij}a_{ji}A_{ji}S_{i\backslash j}^{in}K_{j\backslash i}^{out}.
\end{eqnarray}
Using $\vert w_{\ell+1}\rangle_{ij}=\sum_{kl}\widehat{\mathcal{M}}_{ijkl}\vert w_{\ell}\rangle_{kl}$, we can calculate $\textbf{w}_{2}$. For $\ell=2$,
\begin{eqnarray}
{}_{ij}\langle w_{2}\vert&=&\sum_{kl} {}_{kl}\langle w_{1}\vert \widehat{\mathcal{M}}_{klij}\nonumber \\
&=&\sum_{kl}a_{lk}A_{lk}A_{ji}\delta_{li}\left( 1-\delta_{kj} \right) S_{k\backslash l}^{in}\nonumber \\
&=&A_{ji}\sum_{k\backslash j}a_{ik}A_{ik}S_{k\backslash i}^{in},
\end{eqnarray}
\begin{eqnarray}
\vert w_{2}\rangle_{ij}&=&\sum_{kl} \widehat{\mathcal{M}}_{ijkl}\vert w_{1}\rangle_{kl}\nonumber \\
&=&\sum_{kl}a_{ji}A_{ji}A_{lk}\delta_{jk}\left( 1-\delta_{il} \right) a_{lk}K_{l\backslash k}^{out}\nonumber \\
&=&a_{ji}A_{ji}\sum_{l\backslash i}a_{lj}A_{lj}K_{l\backslash j}^{out}.
\end{eqnarray}
So $\left| \textbf{w}_{2} \right|^{2}$ is:
\begin{eqnarray}\label{lamda2}
\left| \textbf{w}_{2} \right|^{2}&=&\sum_{ij}{}_{ij}\langle w_{2} \vert w_{2}\rangle_{ij}\nonumber \\
&=&\sum_{ijkl}a_{ik}a_{ji}a_{lj}A_{ik}A_{ji}A_{lj}S_{k\backslash i}^{in}K_{l\backslash j}^{out}.
\end{eqnarray}

The above calculation can be extended to any iteration time $\ell$. Specifically, we have
\begin{eqnarray}\label{lamda_l}
\left| \textbf{w}_{\ell} \right|^{2}&=&\sum_{i_{\ell}i_{\ell-1}\cdots i_{1}j_{1}j_{2}\cdots j_{\ell}}\underbrace{a_{i_{\ell-1}i_{\ell}}\cdots a_{i_{1}i_{2}}}_{\ell-1}a_{j_{1}i_{1}}\underbrace{a_{j_{2}j_{1}}\cdots a_{j_{\ell}j_{\ell-1}}}_{\ell-1}\nonumber\\
&\times&\underbrace{A_{i_{\ell-1}i_{\ell}}\cdots A_{i_{1}i_{2}}}_{\ell-1}A_{j_{1}i_{1}}\underbrace{A_{j_{2}j_{1}}\cdots A_{j_{\ell}j_{\ell-1}}}_{\ell-1}\nonumber \\
&\times&S_{i_{\ell}\backslash i_{\ell-1}}^{in}K_{j_{\ell}\backslash j_{\ell-1}}^{out}.
\end{eqnarray}

The collective influence (CI) of the link $i\to j$ is defined as the contribution of the terms containing $a_{ji}A_{ji}$ to $\left| \textbf{w}_{\ell} \right|^{2}$. According to Eq. (\ref{lambda}), a link with a higher CI score will contribute more to the value of $\left| \textbf{w}_{\ell}\right|$, thus $\lambda_{NB}$. If an existing link $i\to j$ is removed from the network, the value of $\left| \textbf{w}_{\ell} \right|^{2}$ will drop by
\begin{eqnarray}\label{E_directed}
\text{CI}_{i\to j}^{\ell}=\sum_{\alpha\beta}S_{\alpha \backslash \alpha^{'}}^{in}K_{\beta\backslash \beta^{'}}^{out}\prod_{(k,l)\in \mathcal{P}_{2\ell-1}^{i\to j}(\alpha,\beta)}a_{lk},
\end{eqnarray}
where $\mathcal{P}_{2\ell-1}^{i\to j}(\alpha,\beta)$ is the set of directed paths of length $2\ell-1$ that cross the link $i\to j$, whose head and tail are the links $\alpha\to\alpha'$ and $\beta'\to\beta$ respectively. In Eq. (\ref{E_directed}), $\alpha$, $\alpha'$, $\beta$ and $\beta'$ run over all possible choices of nodes that satisfy the definition of paths in $\mathcal{P}_{2\ell-1}^{i\to j}(\alpha,\beta)$. In tree structure without loops, the directed path between $\alpha$ and $\beta$ are simply the shortest path connecting them. In locally tree-like networks, we can approximate the directed path using the shortest path from $\alpha$ to $\beta$, because the number of other paths will become negligible as the network size increases \cite{morone2015influence}. In implementation, the shortest paths crossing the link $i\to j$ can be easily found using a breadth-first search algorithm. Notice that, Eq. (\ref{E_directed}) can be generalized for undirected links. Specifically, we treat an undirected edge $i\leftrightarrow j$ as the superposition of two directed links. The CI value of length $\ell$ is defined as $\text{CI}_{ij}^{\ell}=\text{CI}_{i\to j}^{\ell}+\text{CI}_{j\to i}^{\ell}$.

In order to find the optimal set of links, we adopt a heuristic approach following a greedy strategy. Specifically, we adaptively remove the link with the highest CI score until the critical state is reached, i.e., $\lambda_{NB}=1$. After each removal, the CI scores of existing links are recalculated. In practice, we use several techniques to speed up the computation. 1) We limit the path length $\ell$ to a small value to avoid enumerating long paths. Computationally, finding the shortest paths of a fixed length $\ell$ crossing a given link takes $O(1)$ time. As a result, the complexity of computing CI values for all links is $O(M)$. 2) We use a heap structure to find the link with the highest CI score \cite{morone2016collective}, which takes $O(\log M)$ time. 3) After each removal, we only update the CI values of the links within $2\ell$ steps from the removed link, because the CI values of the links outside this scope will not be affected by the removal. This procedure will reduce the computation time of CI update from $O(M)$ to $O(1)$. Combining all these techniques, the overall computational complexity of the algorithm is $O(M\log M)$.

The pseudocode for computing the CI score for an undirected link $i\leftrightarrow j$ is shown below.
\begin{algorithm}[H]
\caption{$\text{CI}(G, i, j, \ell)$}
\label{CIalgorithm}
\begin{algorithmic}[1]
\State $\text{CI}=0$;
\For{each $L_{1}\in [0,2\ell-2]$}
\State $L_{2}=2\ell-2-L_{1}$;
\State $S_{1}=0, S_{2}=0$;
\State $K_{1}=0, K_{2}=0$;
\For{$i_{L_1}\in\partial\text{Ball}(i,L_1)$}
\If{$j\notin\mathcal{D}_{ii_{1}\dots i_{L_{1}}}^{L_{1}}(i,i_{L_1})$}
\State Calculate $S_{i_{L_1} \backslash i_{L_1-1}}^{in}$ and $K_{i_{L_1} \backslash i_{L_1-1}}^{out}$;
\State $P=\prod_{k=0}^{L_{1}-1} a_{i_{k}i_{k+1}}$;
\State $S_{1}+=S_{i_{L_1} \backslash i_{L_1-1}}^{in}\times P$;
\State $K_{1}+=K_{i_{L_1} \backslash i_{L_1-1}}^{out}\times P$;
\EndIf
\EndFor
\For{$j_{L_2}\in\partial\text{Ball}(j,L_2)$}
\If{$i\notin\mathcal{D}_{jj_{1}\dots j_{L_{2}}}^{L_{2}}(j,j_{L_2})$}
\State Calculate $S_{j_{L_2} \backslash j_{L_2-1}}^{in}$ and $K_{j_{L_2} \backslash j_{L_2-1}}^{out}$;
\State $P=\prod_{k=0}^{L_{2}-1} a_{j_{k}j_{k+1}}$;
\State $S_{2}+=S_{j_{L_2} \backslash j_{L_2-1}}^{in}\times P$;
\State $K_{2}+=K_{j_{L_2} \backslash j_{L_2-1}}^{out}\times P$;
\EndIf
\EndFor
\State $\text{CI}+=S_{1}\times K_{2}+S_{2}\times K_{1}$;
\EndFor
\State \textbf{return} $\text{CI}=\text{CI}\times a_{ij}$;
\end{algorithmic}
\end{algorithm}

Here $\partial\text{Ball}(i,L_1)$ is the frontier of the ball centered at node $i$ with radius $L_1$, and $\mathcal{D}_{ii_{1}\dots i_{L_{1}}}^{L_{1}}(i,i_{L_1})$ represents the shortest path of length $L_1$ between node $i$ and $i_{L_1}$. Moreover, $j\notin\mathcal{D}_{ii_{1}\dots i_{L_{1}}}^{L_{1}}(i,i_{L_1})$ means node $j$ is not on the shortest path $i,i_{1},\dots,i_{L_1}$, i.e., $j\notin\{i,i_{1},\dots,i_{L_1}\}$.

\section{\label{sec:5}Competing Methods}

There exist several widely used link ranking methods that can be potentially used in dynamic range maximization \cite{pei2013spreading,lu2016vital,pei2014searching,pei2017theories}. Here we introduce some heuristic methods that is applicable to large-scale networks. For simplicity, in our following analysis, we assume the networks are undirected.

\begin{itemize}

\item Weight-based ranking. To quantify the importance of an undirected link $i\leftrightarrow j$, we define its high weight (HW) score as the sum of links' weights at both ends of the link, i.e., $C^{HW}_{ij}=\sum_{k\in\partial i}a_{ki}+\sum_{k\in\partial j}a_{kj}$, where $\partial i$ is the set of node $i$'s neighbors. In the high weight (HW) ranking, we sort the links by their HW scores in a descending order, and then remove those ranked in the top $q$ fraction. An effective way to improve the performance of HW strategy is to rank the links adaptively. That is, after the removal of each link, the HW scores of the remaining links are recalculated. We refer this method as high weight adaptive (HWA) in the following comparison.

\item Degree-based ranking. In high degree (HD) ranking, the score of each link $i\leftrightarrow j$ is determined by the number of connections at node $i$ and $j$: $C^{HD}_{ij}=\sum_{k\in\partial i}A_{ki}+\sum_{k\in\partial j}A_{kj}$. In the comparison, we also perform the adaptive version of HD ranking (HDA).

\item Eigenvector-based ranking. Many ranking methods make use of the eigenvector corresponding to the largest eigenvalue of the adjacency matrix, which incorporates the information of the global network structure \cite{bonacich1972factoring,brin2012reprint,restrepo2006characterizing}. For instance, the dynamical importance of network links can be characterized by the quantity defined based on eigenvectors \cite{restrepo2006characterizing}. Using a perturbation analysis on the largest eigenvalue, the {\it dynamical importance} of a link $i\to j$ is calculated as
\begin{equation}\label{DI}
I_{ij}=\frac{a_{ij}v_iu_j}{\lambda_{W}v^Tu},
\end{equation}
where $v$ and $u$ denote the right and left eigenvectors of the weighted adjacency matrix $W$. That is, $Wu=\lambda_Wu$ and $v^TW=\lambda_Wv^T$. For undirected networks, the left and right eigenvectors are identical. In the following discussion, we refer this method as Eig ranking.

\end{itemize}

\section{\label{sec:6}Numerical experiments}

\begin{figure}
\includegraphics[width=1\columnwidth]{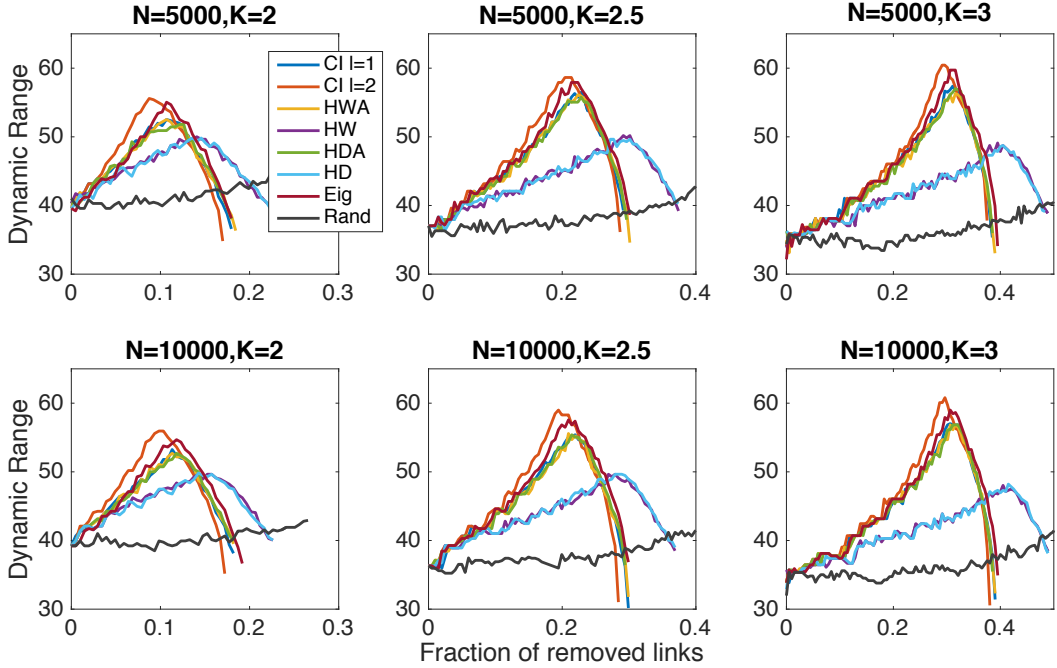}
\caption{\label{ER_Range} For ER networks with size $N=5000$ (top row) and $N=10000$ (bottom row), we use different methods (including CI with $\ell=1$ and $\ell=2$, HWA, HW, HDA, HA, Eig, and Rand) to remove links and display the dynamic range of the remaining networks. Curves for different methods are distinguished by color. The average degree of the network is set as $K=2$ (left column), $K=2.5$ (middle column) and $K=3$ (right column), respectively.}
\end{figure}

To test the performance of CI algorithm, here we examine how the dynamic range evolves after a certain fraction of links are removed. The set of removed links are selected by different methods, including CI algorithm with $\ell=1$ and $2$, HWA, HW, HDA, HD and Eig. For reference, we also use a random selection (Rand) method to pick the links randomly. A better method will result in a smaller fraction of links when the dynamic range is optimized.

We first test on homogeneous ER networks. For ER networks with size $N$ and average degree $K$, we generated the networks by randomly assigning $NK/2$ links among $N$ nodes. Networks with size $N=5000$, $10000$ and average degree $K=2$, $2.5$ and $3$ were used in the simulations. Here we assume the initial $\lambda_{NB}$ is not too far away from 1 so that the critical state can be reached through a small perturbation (i.e., removal of a small fraction of links) on the network structure. To keep the initial $\lambda_{NB}$ close to 1, the link weights were randomly drawn from a uniform distribution $U[0.7,0.8]$. The specific distribution of link weights will not affect the performance of each method. Starting from the initial network structure, links were removed one by one and we calculated the dynamic range at every $0.5\%$ interval. After each link removal, we only kept the giant connected component in the network and discarded the isolated small clusters.

The evolution of the dynamic range for different methods are presented in Fig. \ref{ER_Range}. For all networks, the CI algorithm with $\ell=2$ consistently outperforms other methods by achieving the critical state with a smaller set of links. In addition, the optimal dynamic range for the $\text{CI}_{\ell=2}$ algorithm is higher than those for other competing methods. As the average degree $K$ grows, the critical fraction of links required to optimize the dynamic range increases from less than $10\%$ to around $30\%$. Moreover, the maximal dynamic range also increases from 55 to 60. These results indicate that a better connected ER network needs more links to be removed to reach the critical state, but the optimal dynamic range becomes higher accordingly. The dynamical importance Eig also has a satisfactory result and performs better than other ranking strategies except for the $\text{CI}_{\ell=2}$ algorithm. The performance of the $\text{CI}_{\ell=1}$ algorithm is similar to HWA and HDA, as it only uses local information adaptively like HWA and HDA. With adaptive calculation, HWA and HDA have substantial improvements compared with the original version of HW and HD. Finally, the random selection almost has no effect on improving the dynamic range.

\begin{figure}
\includegraphics[width=1\columnwidth]{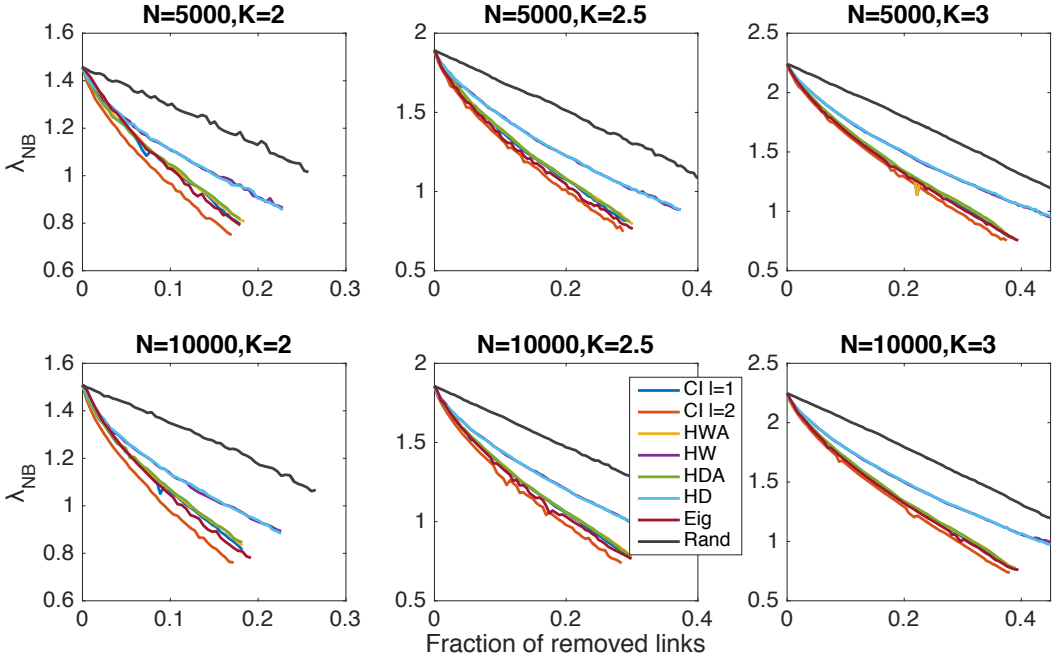}
\caption{\label{ER_Lambda} For ER networks, we present the largest eigenvalue of the WNB matrix $\lambda_{NB}$ after a given fraction of links are removed from the networks using different methods. Parameters of network structure are the same as in Fig. \ref{ER_Range}.}
\end{figure}

To inspect how the largest eigenvalue of the WNB matrix is affected by link removal, we display the value of $\lambda_{NB}$ as a function of removal fraction for different methods in Fig. \ref{ER_Lambda}. For $K=2$ (left column), $\lambda_{NB}$ decreases faster for the $\text{CI}_{\ell=2}$ algorithm than other methods. As the average degree $K$ increases, the $\text{CI}_{\ell=2}$ algorithm remains the best strategy but the difference between Eig, $\text{CI}_{\ell=1}$, HWA and HDA becomes narrower. One possible reason is that the CI algorithm is developed based on locally tree-like structure. There may be more short loops appear as the average degree increases. As a result, the CI algorithm should work more effectively for sparse networks.

\begin{figure}
\includegraphics[width=1\columnwidth]{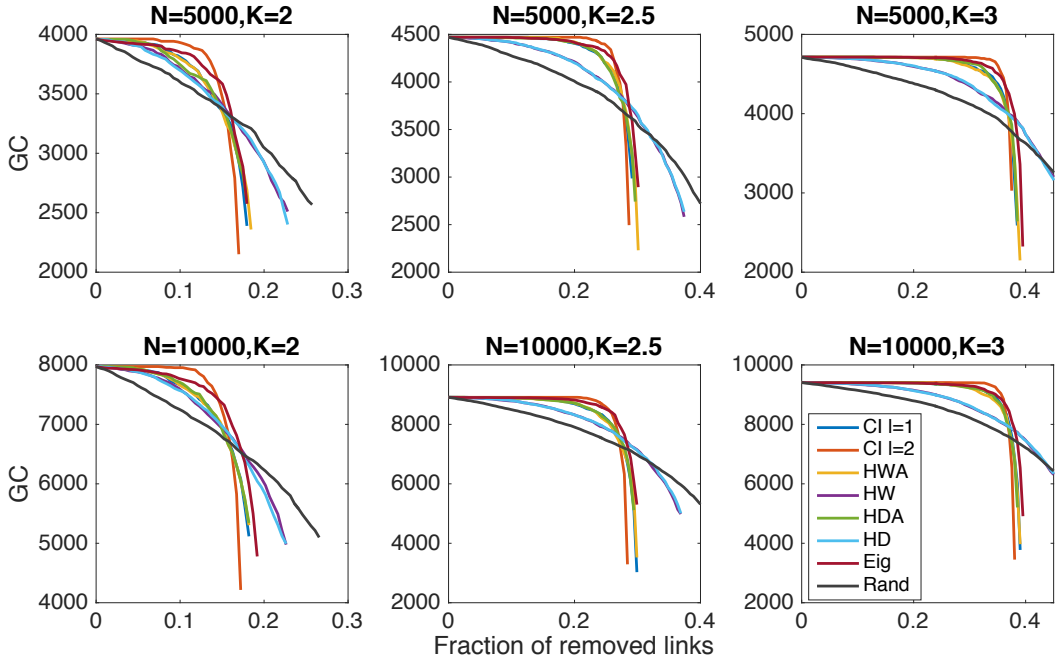}
\caption{\label{ER_GC} For ER networks, we display the size of the giant component (GC) in the remain network with the removal of a certain fraction of links. Different methods are used in selecting removed links. Network structure is the same as in Fig. \ref{ER_Range}.}
\end{figure}

The functionality of excitable networks is not solely determined by the dynamic range. To keep the sensory network working, it is critical to maintain the integrity of the network structure. Here we use the size of the giant component (GC) to quantify the completeness of the network. In particular, the system should reach the critical state before the network gets fragmented by link removal. In Fig. \ref{ER_GC}, we show the decrease of GC size with the removal of links. Before the critical state, GC for the $\text{CI}_{\ell=2}$ algorithm is the largest among all methods, although it drops abruptly afterwards. This implies, the $\text{CI}_{\ell=2}$ algorithm will first select redundant links surrounded by clusters of hubs whose removal will not affect the overall connectivity of the network. As a result, the $\text{CI}_{\ell=2}$ algorithm is suitable for dynamic range maximization, since it will not significantly change the original network structure.

\begin{figure}
\includegraphics[width=1\columnwidth]{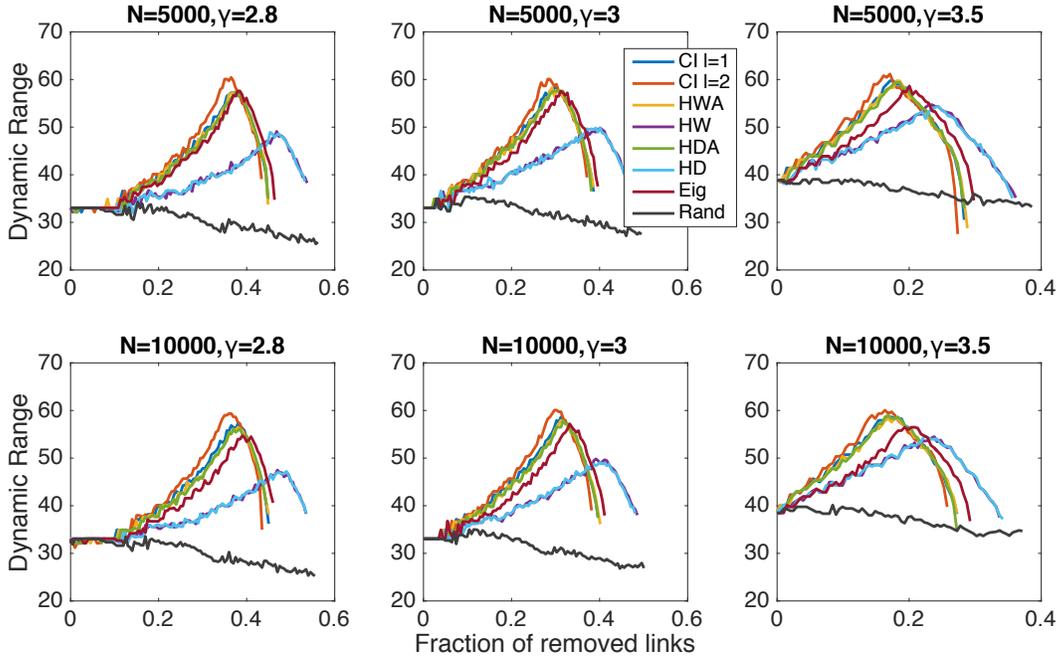}
\caption{\label{SF_Range} We show the dynamic range of scale-free (SF) networks after a certain fraction of links are deleted, using different ranking methods. Networks are generated with size $N=5000$ (top row) and $N=10000$ (bottom row), and power-law exponent $\gamma=2.8$ (left column), $\gamma=3$ (middle column) and $\gamma=3.5$ (right column). }
\end{figure}

Now we compare the performance of different methods on heterogeneous SF networks. We generated undirected SF networks with size $N=5000$, $10000$ and power-law exponent $\gamma=2.8$, $3$ and $3.5$. The link weights are generated from a uniform distribution $U[0.7,0.8]$. We performed the same analysis as for ER networks and present the evolution of dynamic range in Fig. \ref{SF_Range}. Similar with the results on ER networks, the $\text{CI}_{\ell=2}$ algorithm is capable of finding a smaller set of important links. As the network becomes more heterogeneous (the power-law exponent $\gamma$ from 3.5 to 2.8), more fraction of links are required to be removed before the critical state appears. However, there is no noticeable difference in the optimal dynamic range at criticality. For SF networks, the performance of the dynamical importance Eig becomes worse than $\text{CI}_{\ell=1}$, HWA and HDA, which is different from the case of ER networks.

\begin{figure}
\includegraphics[width=1\columnwidth]{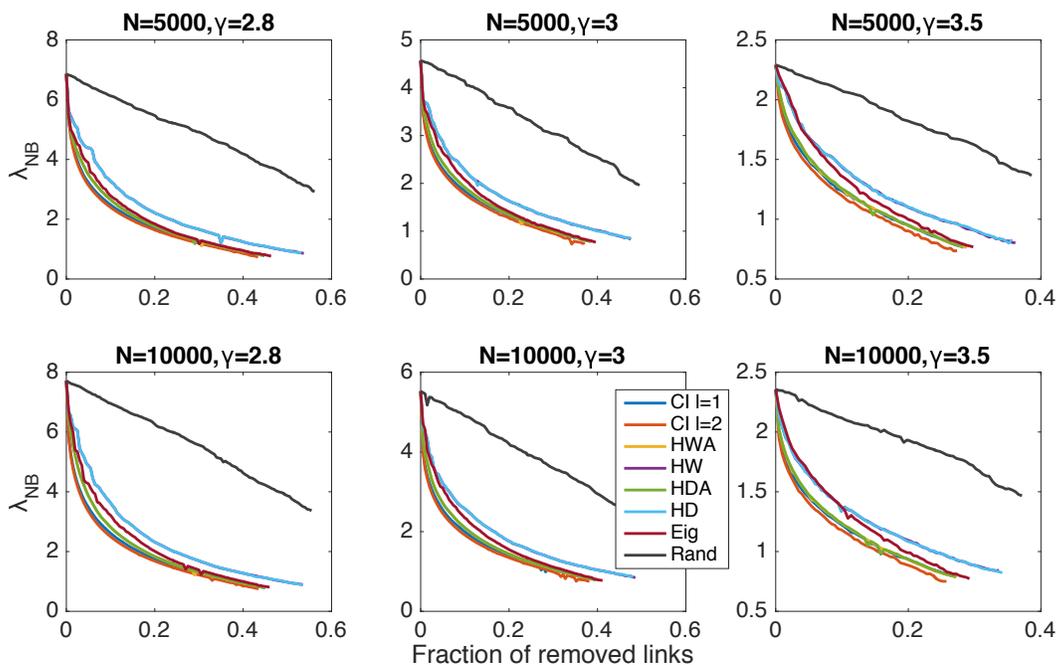}
\caption{\label{SF_Lambda} For SF networks, we show the largest eigenvalue of the WNB matrix $\lambda_{NB}$ after a given fraction of links are removed from the networks.}
\end{figure}

\begin{figure}
\includegraphics[width=1\columnwidth]{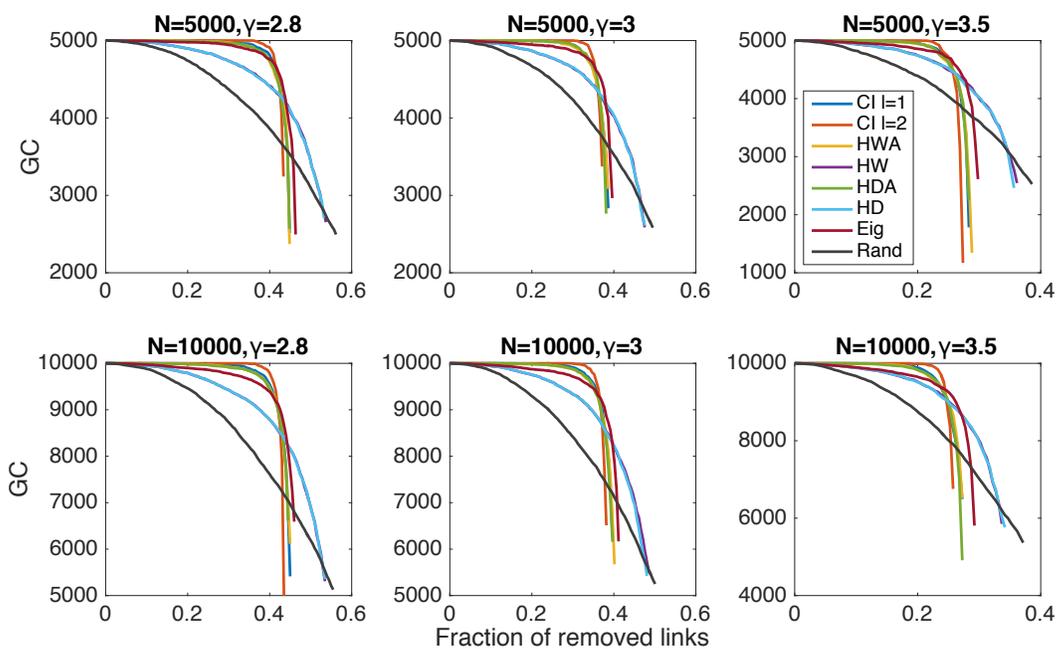}
\caption{\label{SF_GC} For SF networks, we display the size of the giant component (GC) in the remain network with the removal of a certain fraction of links.}
\end{figure}

We display the decrease of the largest eigenvalue of the WNB matrix $\lambda_{NB}$ as a function of the fraction of removed links in Fig. \ref{SF_Lambda}. For SF networks, the difference of the curves for different methods (except random selection) becomes smaller as the network becomes more heterogeneous. This is reasonable because extremely high-degree hubs will emerge in the network, and all methods will tend to select the links between well-connected hubs. Although these curves looks similar, this small difference can result in a noticeable improvement in dynamic range maximization as shown in Fig. \ref{SF_Range}. The change of giant component size for different methods is presented in Fig. \ref{SF_GC}. For all network structure, the $\text{CI}_{\ell=2}$ algorithm has the largest GC size at the critical state. An interesting observation is that, as the network becomes more heterogeneous (the power-law exponent $\gamma$ changes from 3.5 to 2.8), more links are required to be removed to fragment the network. This is different from optimal percolation in which fewer nodes are needed for more heterogenous networks.

\begin{figure}
\includegraphics[width=1\columnwidth]{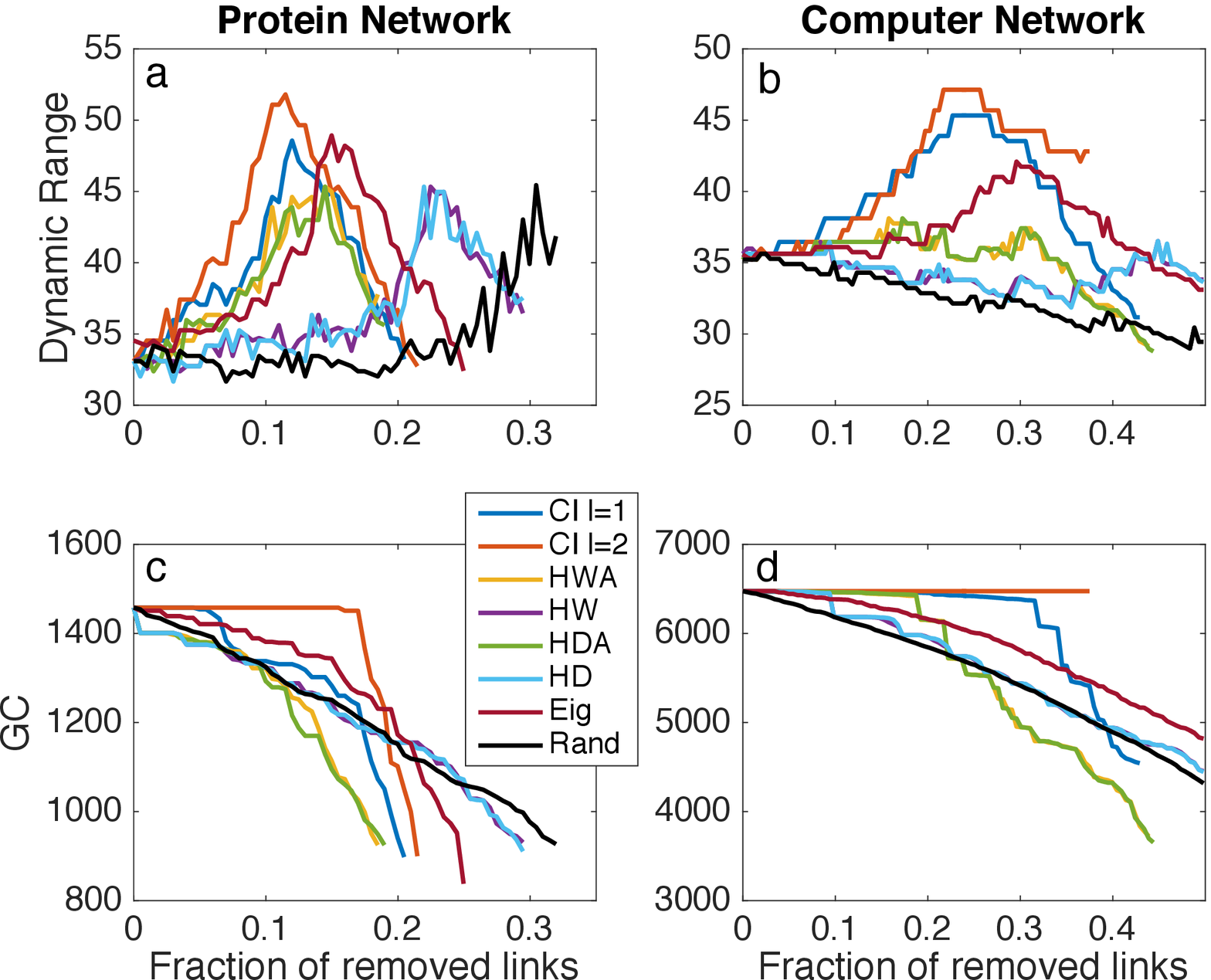}
\caption{\label{Real_Network} For protein network (a) and computer network (b), the dynamic ranges after the removal of a fraction of links are compared for different methods. The changes of GC size for protein network (c) and computer network (d) are also displayed.}
\end{figure}

After the tests on synthetic networks, we turn to more structurally complex real-world networks. Here we select two realistic networks: 1) a yeast protein-protein interaction network (Protein) \cite{coulomb2005gene}, and 2) an Internet autonomous system network (Computer) \cite{leskovec2007graph}. The undirected protein network contains protein interactions contained in yeast. A node represents a protein and an edge represents a metabolic interaction between two proteins. The network consists of 1870 nodes and 2277 links, resulting in an average degree of 2.44. The giant component of the protein network has 1458 nodes. We remove other small clusters in the simulations. The degree sequence follows a power-law distribution with the exponent $\gamma=3.04$. The computer network is the undirected network of autonomous systems of the Internet connected with each other. Nodes are autonomous systems, and edges denote communication. The computer network contains 6474 vertices and 13895 edges, with a power-law degree distribution whose exponent is $\gamma=2.11$. The computer network is fully connected. For both networks, the link weights are randomly drawn from a uniform distribution $U[0.5,0.6]$.

Using different methods, we gradually removed links one by one from the networks. At each $0.5\%$ interval, we calculated the dynamic range of the networks. The evolution of dynamic range is shown in Fig. \ref{Real_Network}(a-b). For protein network in Fig. \ref{Real_Network}(a), removing $11.5\%$ links using $\text{CI}_{\ell=2}$ algorithm can boost the dynamic range from 33.1 to 51.8, a remarkable $56.5\%$ improvement. The CI algorithms with $\ell=2$ and $\ell=1$ outperform other methods, implying that they can handle more complex structure in real-world networks, such as community structure, degree correlation, clustering, etc. For computer network in Fig. \ref{Real_Network}(b), which is more heterogeneous than the protein network, the $\text{CI}_{\ell=2}$ algorithm needs to remove $21.5\%$ links to reach the critical state. This agrees with the result on synthetic SF networks. The dynamic range is enhance from 35.2 to 47.1, a $33.8\%$ improvement. The GC size evolution is presented in Fig. \ref{Real_Network}(c-d). At the critical state, the GC size for the $\text{CI}_{\ell=2}$ algorithm is 1458 (100\%) and 6474 (100\%) for protein network and computer network, respectively. As a result, for real-world networks, the CI algorithm can make the most use of the excitable elements after the adjustment of the network structure. In this case, the $\text{CI}_{\ell=2}$ algorithm preserves all excitable nodes in the networks while others do not. In other competing methods, the Eig ranking outperforms weight-based and degree-based methods, leading to a higher dynamic range and a larger GC size at the critical state.

\section{\label{sec:7}Conclusions}

In this study, we explored the strategy to maximize the dynamic range of a given excitable network by removing a minimal number of links. We used a message passing process to describe the activation propagation in excitable networks and found that the dynamic range is optimized at the critical state, where the largest eigenvalue of the WNB matrix is exactly one ($\lambda_{NB}=1$). To quantify the collective influence (CI) of each single link on $\lambda_{NB}$, we analyzed the message passing equations through linearization and approximated the largest eigenvalue using power method iteration. Based on the CI scores of links, we proposed a greedy algorithm that adaptively remove edges with the largest CI value. Simulations on both synthetic and real-world networks indicate that, compared with other widely used heuristics, the CI algorithm can find a smaller set of links whose removal will perturb the network to the critical state. Moreover, at the critical state, the optimal dynamic range for CI algorithm is higher than other methods. Meanwhile, CI algorithm maintains a larger giant component after the removal of links, which preserves the majority of excitable elements in the network. The proposed method provides a practical way to enhance the dynamic range of excitable networks through a minimal control of the network structure, and can be generalized to quantify the dynamical importance of links in other dynamical processes.

\begin{acknowledgments}
This work is partially supported by National Natural Science Foundation of China (11290143), China Postdoctoral Science Foundation (2015M581324), the Fundamental Research Funds for the Central Universities (DUT16LK39), and the Fundamental Research of Civil Aircraft (No. MJ-F-2012-04).
\end{acknowledgments}

%\nocite{*}
%\bibliography{}

\bibliography{}

\end{document}